\documentclass[aps,pra,a4paper,11pt,twocolumn,superscriptaddress,longbibliography,reprint]{revtex4-1}

\usepackage[english]{babel}
\usepackage[utf8]{inputenc}
\usepackage[colorlinks=true,citecolor=blue]{hyperref}
\usepackage{graphicx,color}
\usepackage{amsmath,amssymb,bm}
\usepackage{epsf,epsfig}
\usepackage[shortlabels,inline]{enumitem}

\newcommand{\ketbra}[2]{|{#1}\rangle\langle{#2}|}
\newcommand{\ket}[1]{|{#1}\rangle}

\begin{document}
\title{Quantum Photovoltaic Cells Driven by Photon Pulses}
\author{Sangchul Oh}
\email[]{soh@hbku.edu.qa}
\affiliation{Qatar Environment and Energy Research Institute,
Hamad Bin Khalifa University, Qatar Foundation, P.O.Box 5825, Doha, Qatar}
\author{Jungjun Park}
\affiliation{Korea Institute for Advanced Study,
85 Hoegiro, Dongdaemun-gu Seoul 02455, Republic of Korea}
\author{Hyunchul Nha}
\affiliation{Texas A\&M University at Qatar, P.O.Box 23874, Education City, Doha, Qatar}
\date{\today}

\begin{abstract}
We investigate the quantum thermodynamics of two quantum systems, a two-level system and a four-level quantum 
photocell, each driven by photon pulses as a quantum heat engine. We set these systems to be in thermal 
contact only with a cold reservoir while the heat (energy) source, conventionally given from a hot thermal 
reservoir, is supplied by a sequence of photon pulses. The dynamics of each system is governed by a coherent 
interaction due to photon pulses in terms of the Jaynes-Cummings Hamiltonian together with the system-bath 
interaction described by the Lindblad master equation. We calculate the thermodynamic quantities for 
the two-level system and the quantum photocell including the change in system energy, power delivered by 
photon pulses, power output to an external load, heat dissipated to a cold bath, and entropy production. 
We thereby demonstrate how a quantum photocell in the cold bath can operate as a continuum quantum heat 
engine with the sequence of photon pulses continuously applied. We specifically introduce the power efficiency
of the quantum photocell in terms of the ratio of output power delivered to an external load with current 
and voltage to the input power delivered by the photon pulse. Our study indicates a possibility that a quantum 
system driven by external fields can act as an efficient quantum heat engine under non-equilibrium thermodynamics.
\end{abstract}
\maketitle

\section{Introduction\label{Introduction}}

Thermodynamics deals with the evolution of systems, usually in contact with reservoirs, describing the dynamics 
under universal laws independent of microscopic details. Among its four laws, the second law dictates the total 
entropy of a closed system can never decrease over time and that the closed system spontaneously 
evolves toward the state with maximum entropy. One of the possible statements about the second law of thermodynamics 
is to set the upper bound on the efficiency of heat engines. Heat engines convert heat energy, which typically flows 
from a hot source to a cold sink, to mechanical energy or chemical energy. The efficiency of energy conversion is 
defined by the ratio of the work output to the amount of heat energy input. The ultimate efficiency of the heat engine 
is known in equilibrium thermodynamics to be determined only by temperatures  of hot and cold heat baths, $T_h$ 
and $T_c$, respectively, i.e. $\eta = 1 - T_c/T_h$, the so-called the Carnot limit.

Photovoltaic cells (or solar cells) and photosynthesis, just like classical heat engines, convert photon energy 
from the sun into electric energy and chemical energy, respectively. The upper limit of the efficiency of 
{\it p-n} junction solar cells with an energy bandgap is known as the Shockley-Queisser limit~\cite{Shockley1961}.
The key assumptions in deriving the Shockley-Queisser limit are (i) photons with energies less than the bandgap 
are not utilized, (ii) a photon with energy greater than the bandgap produces only one electron-hole pair and 
(iii) only the radiative recombination of electron-hole pairs is considered. While the non-radiative loss may be 
minimized by the manufacturing technology, radiative recombination is the intrinsic energy loss governed by 
the law of physics. Assuming the sun and the solar cell are described as black-bodies with temperature 
$T_s = 6000\;{\rm K}$ and $T_c = 300\;{\rm K}$, respectively, the maximum efficiency is about 30\% for 
a solar cell with a bandgap of $1.137 {\;\rm eV}$~\cite{Shockley1961}. Shockley and Queisser~\cite{Shockley1961} 
also showed that the maximum efficiency of a single {\it p-n} junction solar cell would be approximately 
44\% around 1.137{\;\rm eV} if there is no radiative recombination loss.

Recently, many theoretical studies have suggested that noise-induced quantum coherence~\cite{Scully2010,Scully2011},
Fano-induced coherence~\cite{Svidzinsky2011} or delocalized quantum states of interacting 
dipoles~\cite{Creatore2013,Oh2015,Fruchtman2016,Higgins2017} can reduce the radiative recombination loss of 
a solar cell, thus enhancing the efficiency of solar cells. The same idea is applied to the photosynthetic 
complex~\cite{Dorfman2013,Killoran2015,Yamada2015,Stones2017}.  Most of these studies employ the donor-acceptor 
quantum photocell model where the donor is in thermal equilibrium with a hot bath, i.e., the sun at 
$5800 {\; \rm K}$ and the acceptor is at room temperature. The photocell operating while continuously contacting
both heat reservoirs is called  a continuous quantum heat engine~\cite{Kosloff2014,Uzdin2015}.
A typical example of the continuous quantum heat engine is Scovil and Schulz-Dubois' three-level masers 
whose efficiency achieves the Carnot efficiency~\cite{Scovil59}. By solving the master equation for 
the quantum photocell, it was shown that the noise-induce quantum coherence or the dark state of the donor may 
enhance the power output. However, it was not clear whether the efficiency of the photocell could be enhanced by 
the quantum effect.  Some works assumed that the mean photon number of the hot thermal bath 
could be $\bar{n}= 60000$, but the mean photon number of the sun at energy $1.8{\; \rm eV}$ as a black body 
at $5800{\;\rm K}$ is only about $\bar{n} = 0.037$~\cite{Oh2019,Tomasi2020}. To address this issue, Ref.~\cite{Oh2019}
introduced the pumping term and showed that the dark state could enhance power output but not its efficiency.

In this paper, we explore another form of quantum heat engine. We consider two quantum systems, a two-level system 
and a donor-acceptor quantum photocell, and investigate their quantum dynamics under coherent driving and 
system-bath interaction. Each quantum system is in thermal contact only with a cold reservoir but not with a hot 
reservoir. Instead, they are driven by a sequence of photon pulses that supply input energy to the systems, 
which is conventionally done by a hot reservoir. The photon pulses represent the stream of energy source to 
the system and may thus remove the unrealistic assumption of high mean photon number of the sun by the previous works. 
We solve the time-dependent Markov Lindblad master equation and investigate the thermodynamic quantities such 
as the change in energy, the heat dissipation to the cold bath, the power delivered by the photon pulse and 
the entropy generation. Specifically we introduce the power efficiency of the quantum photocell in terms of 
the ratio of output power delivered to an external load to input power delivered by the photon pulses. 

This paper is organized as follows. In Sec.~\ref{Sec:Q_thermo}, we review briefly the quantum dynamics 
and quantum thermodynamics of an open quantum system based on the master equation approach. 
In Sec.~\ref{Sec:Two-level}, we examine a two-level system in a cold bath driven by photon pulses, and 
examine how the energy, heat current, and entropy change. In Sec.~\ref{Sec:Photocell} a quantum 
photovoltaic cell with donor and acceptor driven by repeated photon pulses is considered. We calculate 
the quantum thermodynamic quantities and the power output by the sequence of the photon pulses together 
with engine efficiency.  Finally, in Sec.~\ref{Sec:Summary} we summarize our results with some discussion.

\section{Quantum thermodynamics of open quantum systems}
\label{Sec:Q_thermo}

We start with a brief review of quantum thermodynamics of an open quantum system which exchanges energy and 
entropy with its environment. The equations presented in this section will be applied to those examples 
in next two sections~\ref{Sec:Two-level} and \ref{Sec:Photocell}.
As usual, we assume the Born-Markov approximations: a weak interaction between an open quantum system and 
the environment, and the extremely short correlation time of the environment, i.e., no memory effect. 
The density operator $\rho(t)$ of the quantum system with a slowly varying time-dependent Hamiltonian 
obeys the Lindblad-Gorini-Kossakowski-Sudarshan (LGKS) master 
equation~\cite{Lindblad1976,Gorini1976,Alicki1984,Breuer2002,Rivas2012}
\begin{equation}
\frac{d}{dt}\rho(t) ={\cal L}[\rho(t)]
                    = -\frac{i}{\hbar}[H_S(t),\rho(t)] + {\cal D}[\rho(t)] \,,
\label{Eq:Lindblad1}
\end{equation}
where  $H_S(t) = H_0 + H_1(t)$ is the Hamiltonian of the system. Here $H_0$ represents 
a time-independent unperturbed Hamiltonian and $H_1(t)$ an external time-dependent perturbation. 
The decoherence and the dissipation of the open quantum system due to environmental interaction are 
described by the non-unitary  operator
\begin{equation}
{\cal D}[\rho] = \sum_k \left(2L_k\rho L_k^\dag - L^\dag_k L\rho -\rho L^\dag_k L_k \right) \,,
\end{equation}
where $L_k$ are the Lindblad operators determined according to the type of interaction.

From the solution $\rho(t)$ of Eq.~(\ref{Eq:Lindblad1}), one can calculate the quantum thermodynamic 
quantities.  The first law of classical thermodynamics states the energy conservation, 
$dE = \delta Q + \delta W$. The time-dependent internal energy of the system is given by 
$E(t) = {\rm tr}\left\{\rho(t) H_S(t)\right\}$. Its derivative with respect to time gives rise to 
the first law of quantum thermodynamics~\cite{Spohn1978,Spohn1978b,Alicki1979}
\begin{equation}
\frac{d}{dt}E(t) = \dot{Q}(t) +\dot{W}(t) = J(t) +P(t) \,.
\label{1st_Q_thermo}
\end{equation}
Here $J(t)$ is the heat current from the environment into the system
\begin{equation}
J(t) \equiv \dot{Q}(t) = {\rm tr}\left(\frac{d\rho(t)}{dt} H_S(t)\right) \,,
\end{equation}
and $P(t)$ is the power delivered to system by external forces,
\begin{equation}
P(t)\equiv \dot{W}(t) = {\rm tr}\left(\rho(t)\frac{dH_S}{dt}\right)\,.
\end{equation}
Since $H_0$ is the time-independent Hamiltonian, the power can be written as 
$P(t)= {\rm tr}\left(\rho(t)\frac{dH_1(t)}{dt}\right)$. The change in energy of the system for 
finite time can be obtained by integrating the heat current and the power as
\begin{equation}
\Delta E(t) = Q(t) +W(t) = \int_0^tJ(s)\,ds + \int_0^t P(s)\,ds\,.
\end{equation}

The second law of thermodynamics describes the irreversibility of dynamics, where the entropy plays 
a key role.  The von Neumann entropy $S(t)$ of the system in the state $\rho(t)$ is given by 
\begin{equation}
S(t) = -{\rm tr}\left\{\rho(t)\log \rho(t)\right\}\,.
\end{equation}
The thermodynamic entropy ${\cal S}$ is written as ${\cal S} = k_BS(t)$ with the Boltzmann constant 
$k_B$.  The net change in the entropy $dS_{\rm net} $(entropy production) of the whole system+reservoir 
can be written in terms of the entropy change of the system, $dS$, and the entropy flow due to heat 
from environment to system, $dS_e$, as 
\begin{equation}
dS_{\rm net} = dS - dS_e\,.
\end{equation}
The change in the entropy of the system, $dS$, over time is written as
\begin{subequations}
\begin{align}
\frac{dS}{dt} &= -{\rm tr}\left\{\dot{\rho}(t)\log \rho(t)\right\} \\[5pt]
           &= -{\rm tr}\left\{{\cal L}[\rho(t)] \log \rho(t)\right\} \,,
\end{align}
\end{subequations}
where ${\rm tr}\left\{\dot{\rho}(t)\right\} = 0$ and the quantum Markov master equation~(\ref{Eq:Lindblad1}) 
are used. The entropy flow $dS_e$ per unit time from the environment into the system is written as
\begin{subequations}
\begin{align}
\frac{dS_e}{dt}&\equiv J_S= \beta\dot{Q}(t) \\
               &= \beta\,{\rm tr}\left\{\dot{\rho} H_S(t)\right\} 
                = \beta\,{\rm tr}\left\{{\cal L}[\rho(t)]\, H_S(t)\right\} \,,
\end{align}
\end{subequations}
where $\beta = {1}/{k_BT}$ and $T$ is the temperature of the environment.
The net entropy production rate $\sigma(t)$ of the system is given by
\begin{equation}
\sigma(t) = \frac{dS_{\rm net}}{dt} =  \dot{S}(t) -\beta \dot{Q}(t) \ge 0\,,
\label{Eq:entropy_production}
\end{equation}
where $\sigma(t) \ge 0$ comes from the Spohn inequality~\cite{Spohn1978,Spohn1978b,Alicki1979,Das2018}.
Eq.~(\ref{Eq:entropy_production}) may be written as
\begin{equation}
\sigma\equiv -\frac{d}{dt}S(\rho(t)\parallel \rho_{\rm ss})\ge 0\,,
\end{equation}
where $S(\rho(t) \,\|\, \rho_{\rm ss})\equiv {\rm tr}[\rho(t) (\log\rho(t) -\log\rho_{\rm ss})]$ is 
the relative entropy of $\rho(t)$ with respect to the stationary state $\rho_{\rm ss}$,
for example, the canonical state of the system, $\rho_{\rm ss} =\rho_\beta= e^{-\beta H_s(t)}/Z$.
This is called the second law of non-equilibrium quantum thermodynamics in the weak coupling limit.

\section{A Two-level System Driven by Photon Pulses}
\label{Sec:Two-level}

\begin{figure}[t]
\includegraphics[width=0.45\textwidth]{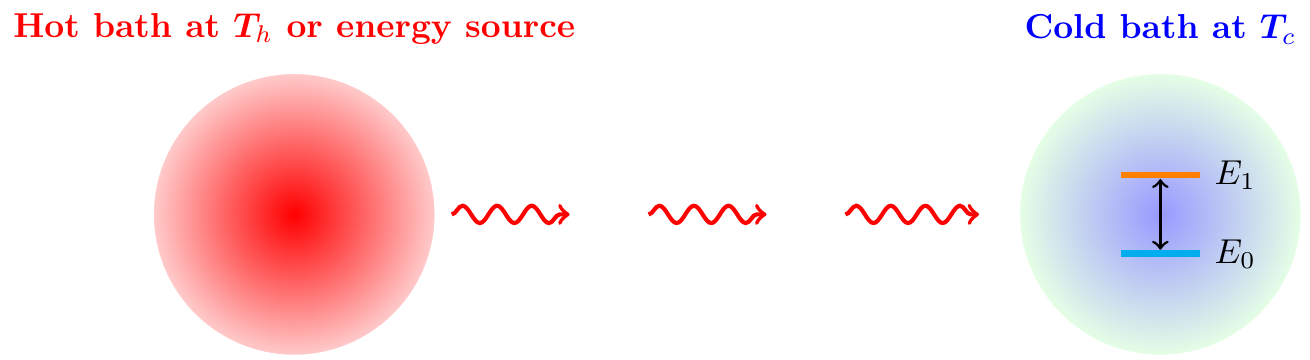}
\caption{A two-level system with energy levels $E_0$ and $E_1$ in contact with a cold thermal bath at $T_c$ is 
driven by Gaussian photon pulses serving as an energy source in our work. } 
\label{Fig:2level_Qsystem}
\end{figure}

As an application of quantum thermodynamics of open quantum systems presented in Sec.~\ref{Sec:Q_thermo}, 
we first consider a two-level quantum system which is in contact with a cold bath at temperature $T_c$ 
and driven by repeated photon pulses, as depicted in Fig.~\ref{Fig:2level_Qsystem}. The hot thermal bath 
supplying energy does not have a direct contact with the quantum system. Its role is here replaced by 
a sequence of photon pulses to the two-level system. We examine how the two-level system absorbs and dissipates 
energy and generates entropy during this process in order to gain insight into the nonequlibrium dynamics due to 
photon pulses.

The unperturbed Hamiltonian of the two-level system with energy levels 
$E_0$ and $E_1$ may be written as
\begin{align}
H_0 = -\frac{\hbar\omega_0}{2}\sigma_z,
\label{Eq:two_level_H0}
\end{align}
where $\omega_0 = (E_1 -E_0)/\hbar$ and $\sigma_z = \ketbra{0}{0} -\ketbra{1}{1}$.
The interaction between a two-level system and incoming photon pulses is described 
by the Jaynes-Cummings Hamiltonian~\cite{Jaynes1963}
\begin{equation}
H_1(t) = i\hbar\left[\;  g^*(t)\sqrt{\gamma}\,\sigma_{-} - g(t)\sqrt{\gamma}\,\sigma_+ \;\right]\,,
\label{Eq:JC_Hamiltonian}
\end{equation}
where $\sigma_{+} = \ketbra{1}{0}$ and $\sigma_{-} = \ketbra{0}{1}$ are the raising and the lowering operators,
respectively. We set $E_1 - E_0 = 1\; {\rm eV}$. 
Here $\gamma$ is the Weisskopf-Wigner spontaneous decay rate 
\begin{align}
\gamma = \frac{1}{4\pi\epsilon_0}\frac{4\omega_0^3 d_{01}^2}{3\hbar c^3}\,,
\label{Eq:WW_constant}
\end{align}
where $d_{01}$ is the transition dipole moment between the two states $\ket{0}$ and $\ket{1}$. 
A typical value of $\gamma$ for an atom or a quantum dot for visible light emission is in the order
of nano seconds corresponding to $\mu{\rm eV}$, while the energy at visible frequencies is about eV, 
i.e., femto seconds. As a numerical calculation becomes demanding with a big difference between these 
time-scales,  we use for our study the values of parameters as listed in Table~\ref{Table1}. We consider 
the photon pulses given at peak times $t_i$ as $g(t)=\alpha\sum_i\xi(t;t_i)$ with coherent states having 
average photon number $\langle n \rangle = |\alpha|^2$ and a Gaussian pulse 
shape~\cite{Loudon1973,Wang2011,Chan2018}
\begin{align}
\xi(t;t_i) \equiv
   \left(\frac{\Omega^2}{2\pi}\right)^{1/4}\,\exp\left[-\frac{\Omega^2 (t-t_i)^2}{4} -i\omega_0 t\right]\,, 
\label{Eq:Gaussian_pulse}
\end{align}
Here $1/\Omega$ is the pulse bandwidth.

Under the Born-Markov approximation, the interaction of a two-level system and the thermal photon bath 
is recast to the dissipative operator ${\cal D}$ acting on the density matrix of 
the system~\cite{Carmichael,Breuer2002}
\begin{align}
{\cal D}_C[\rho]
&= \frac{\gamma}{2}\bigl(\bar{n}_c + 1\bigr)\bigl(2\sigma_{-}\rho\sigma_{+} -\sigma_{+}\sigma_{-}\rho +
\rho\sigma_{+}\sigma_{-}\bigr) \nonumber\\
&+\frac{\gamma}{2}\bar{n}_c
\left(2\sigma_{+}\rho\sigma_{-} -\sigma_{-}\sigma_{+}\rho + \rho\sigma_{-}\sigma_{+}\right)\,.
\label{Eq:two_level_Dc}
\end{align}
Here $\bar{n}_c$ is the mean photon number of the cold bath at the frequency $\omega_0$ in thermal equilibrium 
of temperature $T_c$ 
\begin{align}
\bar{n}_c = \frac{1}{e^{\hbar\omega_0/k_BT_c} -1}\,.
\end{align}
As noted in Refs.~\cite{Carmichael,Mandel1995}, at optical frequencies and room temperature, the mean photon 
number $\bar{n}$ is very small and negligible while it has a finite value at microwave frequencies and 
the room temperature. For example, with $\hbar\omega_0 = \;{\rm eV}$ and $T_c =300\;{\rm K}$
one obtains $\bar{n}\approx 6.5\times10^{-31}$. 
At the optical frequencies, $\hbar\omega_0 = 1.8\;{\rm eV}$ and the temperature of the sun as 
a black body, $T_s = 5800\;{\rm K}$, the mean photon number is $\bar{n}\approx 0.0317$~\cite{Mandel1995,Oh2019}.
With this in mind, Eq.~(\ref{Eq:two_level_Dc}) reduces to 
\begin{align}
{\cal D}_C[\rho]
\approx \frac{\gamma}{2}\bigl(2\sigma_{-}\rho\sigma_{+} -\sigma_{+}\sigma_{-}\rho +
\rho\sigma_{+}\sigma_{-}\bigr) \,.
\end{align}

\begin{table*}[t]
\setlength{\tabcolsep}{12pt}
\begin{center}
\caption{Typical parameters used in this work}
\begin{tabular}{rl}
\hline\hline\\[-5pt]
Energy gap of the two-level system & $E_1-E_0 = \hbar\omega_0 = 1.0$\;eV \\[2pt]
Energy gap of the donor of the quantum photocell & $E_1-E_0 = \hbar\omega_0 = 1.8$\;eV \\[2pt]
Energy gap of the acceptor of the quantum photocell & $E_2 -E_3 = 1.6$\;eV\\[2pt]
Weisskopf-Winger constant         & $\hbar\gamma = 1.24\;\mu{\rm eV}\;\sim 1.0\; {\rm meV}$\\[2pt]
Phonon decay constant             & $\hbar\gamma_{12} = \hbar\gamma_{30} =12\sim 120\;{\rm meV}$\\[2pt]
Photon number of a pulse & $\langle n\rangle =|\alpha|^2 = 1\;{\rm or}\; 10$\\[2pt]
Temperature of the cold bath     & $T_c$ = 300\;K \\[2pt]
Width of a Gaussian pulse        & $\Omega = \omega_0/4\pi$\\[2pt]
\hline\hline
\end{tabular}
\label{Table1}
\end{center}
\end{table*}

\begin{figure}[ht]
\centering{ \includegraphics[width=0.48\textwidth]{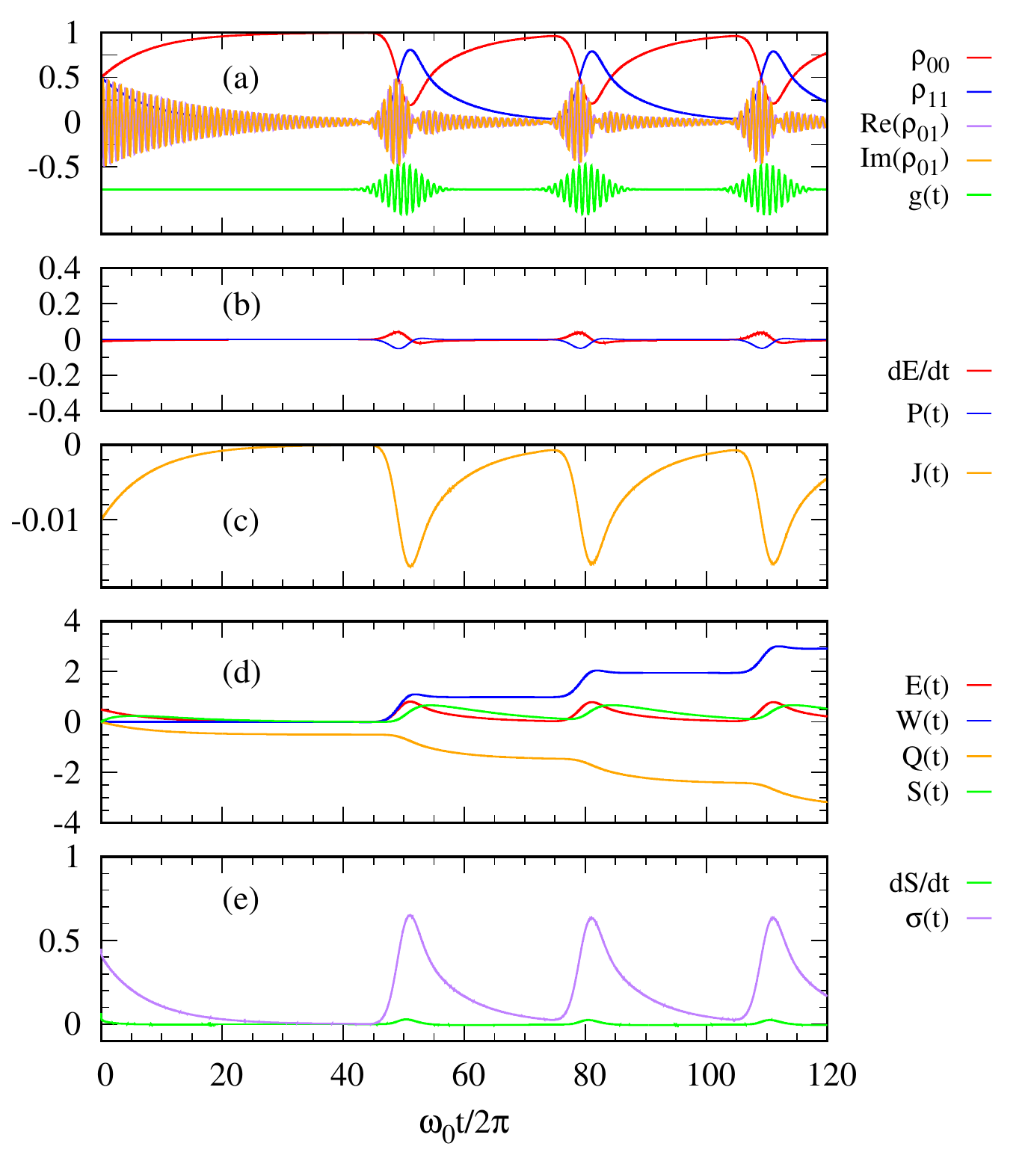}}
\caption{(a) The density matrix elements of the two-level system and the sequence of Gaussian photon 
pulses $g(t)$ are plotted over time. (b) The rate of energy change $\frac{d E(t)}{dt}$ and the power $P(t)$, 
(c) the heat current $J(t)$ are calculated as functions of time. (d) The energy $E(t)$, the work $W(t)$, 
the heat transfer $Q(t)$ and the system entropy $S(t)$ are calculated as functions of time. (e) The rate of 
system entropy change and the entropy production are calculated over time. The parameters are taken as 
$\langle n\rangle  = 1$, $\gamma = 10^{-2}\omega_0$, $\Omega = \omega_0/4\pi$, and 
$\hbar\omega_0 = 1\;{\rm eV}$.
}
\label{Fig2}
\end{figure}
\begin{figure}[ht]
\begin{center}
\includegraphics[width=0.48\textwidth]{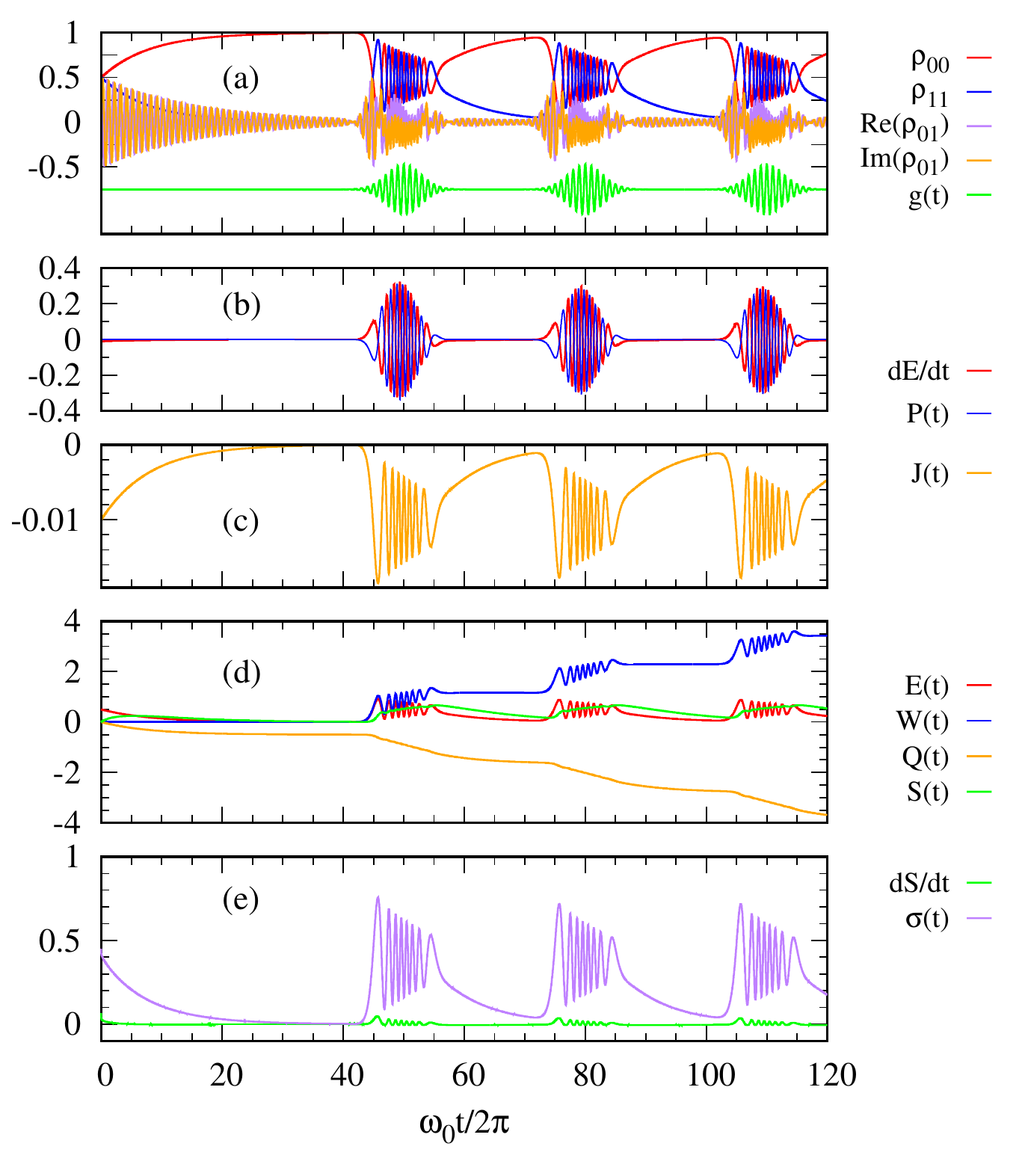}
\end{center}
\caption{(a) The density matrix elements of the two-level system and the sequence of Gaussian photon 
pulses $g(t)$ are plotted over time. (b) The rate of energy change $\frac{d E(t)}{dt}$ and the power $P(t)$, 
(c) the heat current $J(t)$ are calculated as functions of time. (d) The energy $E(t)$, the work $W(t)$, 
the heat transfer $Q(t)$ and the system entropy $S(t)$ are calculated as functions of time. 
(e) The rate of system entropy change and the entropy production are calculated over time.
The parameters are $\langle n\rangle  = 10$, $\gamma = 10^{-2}\omega_0$, $\Omega = \omega_0/4\pi$,
and $\hbar\omega_0 = 1\;{\rm eV}$.
}
\label{Fig3}
\end{figure}

With Eqs.~(\ref{Eq:two_level_H0}), (\ref{Eq:two_level_Dc}), and (\ref{Eq:JC_Hamiltonian}), the Lindblad 
equation for the two-level system, in contact with the cold thermal bath and driven by a Gaussian photon pulse, 
is given by 
\begin{equation}
\frac{d}{dt}\rho(t) = -\frac{i}{\hbar}[H_0 + H_1(t),\rho(t)] + {\cal D}_C[\rho(t)] \,.
\label{Eq:Lindblad}
\end{equation}
The quantum dynamics and the quantum thermodynamics of the two level system are investigated by solving 
Eq.~(\ref{Eq:Lindblad}) numerically using the Runge-Kutta method. 
The parameters used in numerical 
simulation are shown in Table~\ref{Table1}. 

Fig.~\ref{Fig2} and Fig.~\ref{Fig3} describe the thermodynamic quantities of the system when a sequence of Gaussian 
photon pulses are applied at regular interval ($g(t)$: green curves in (a)) with $\langle n\rangle  = 1$ 
and $\langle n\rangle  = 10$, respectively. 
Fig.~\ref{Fig2} (a) shows the time-evolution of the density matrix elements 
and the sequence of Gaussian photon pulses, which is first applied around the peak time $\omega_0t/2\pi = 50$ 
with $\langle n\rangle  = 1$. The initial state of the two level system is assumed to be in a superposed state 
$\ket{\psi(0)} = \frac{1}{\sqrt{2}}\left(\ket{0} + \ket{1}\right)$. 
As shown in Fig.~\ref{Fig2} (a), the superposed 
state decays to the ground state, i.e., the system becomes in thermal equilibrium with the cold thermal bath before 
the photon pulse is applied. When the Gaussian photon pulse is first applied around $\omega_0t/2\pi =50$, the system 
gets excited and then becomes decayed into the ground state after the pulse is gone. This process is repeated 
according to each Gaussian pulse.

\begin{figure}[t]
\centering{
\includegraphics[width=0.48\textwidth]{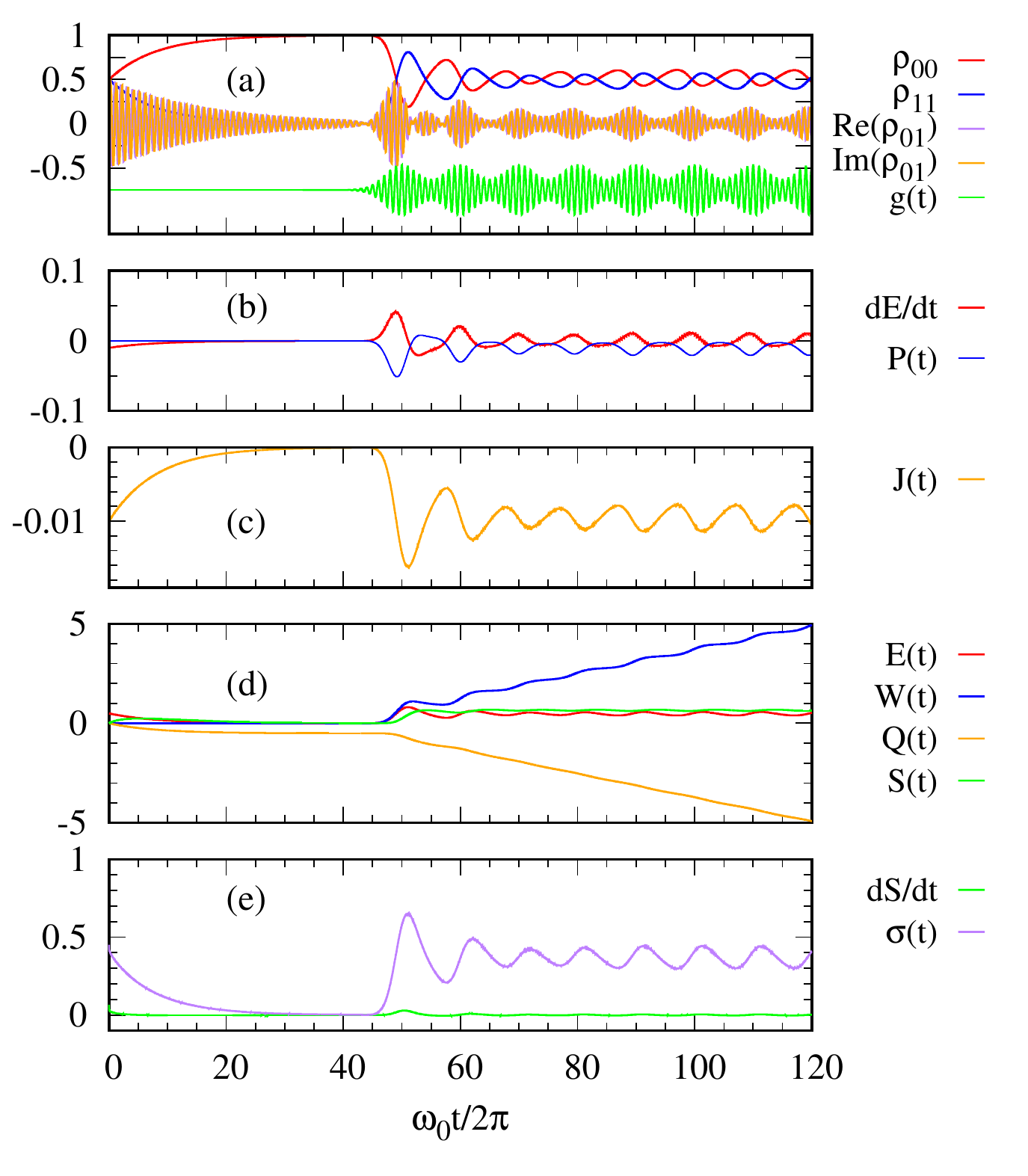}
}
\caption{(a) The density matrix elements of the two-level system and the sequence of Gaussian photon 
pulses $g(t)$,  (b) rate of energy change $\frac{d E(t)}{dt}$ and power $P(t)$, (c) heat current $J(t)$,  
(d) energy $E(t)$, work $W(t)$, heat $Q(t)$ and system entropy $S(t)$, (e) rate of system entropy change and 
entropy production, all as functions of time. This describes the case of regularly spaced sequence of photon pulses.
Parameters: $\langle n\rangle  = 1$, $\gamma = 10^{-2}\omega_0$, $\Omega = \omega_0/4\pi$,
and $\hbar\omega_0 = 1{\;rm eV}$.}
\label{Fig4}
\end{figure}

\begin{figure}[t]
\centering{
\includegraphics[width=0.48\textwidth]{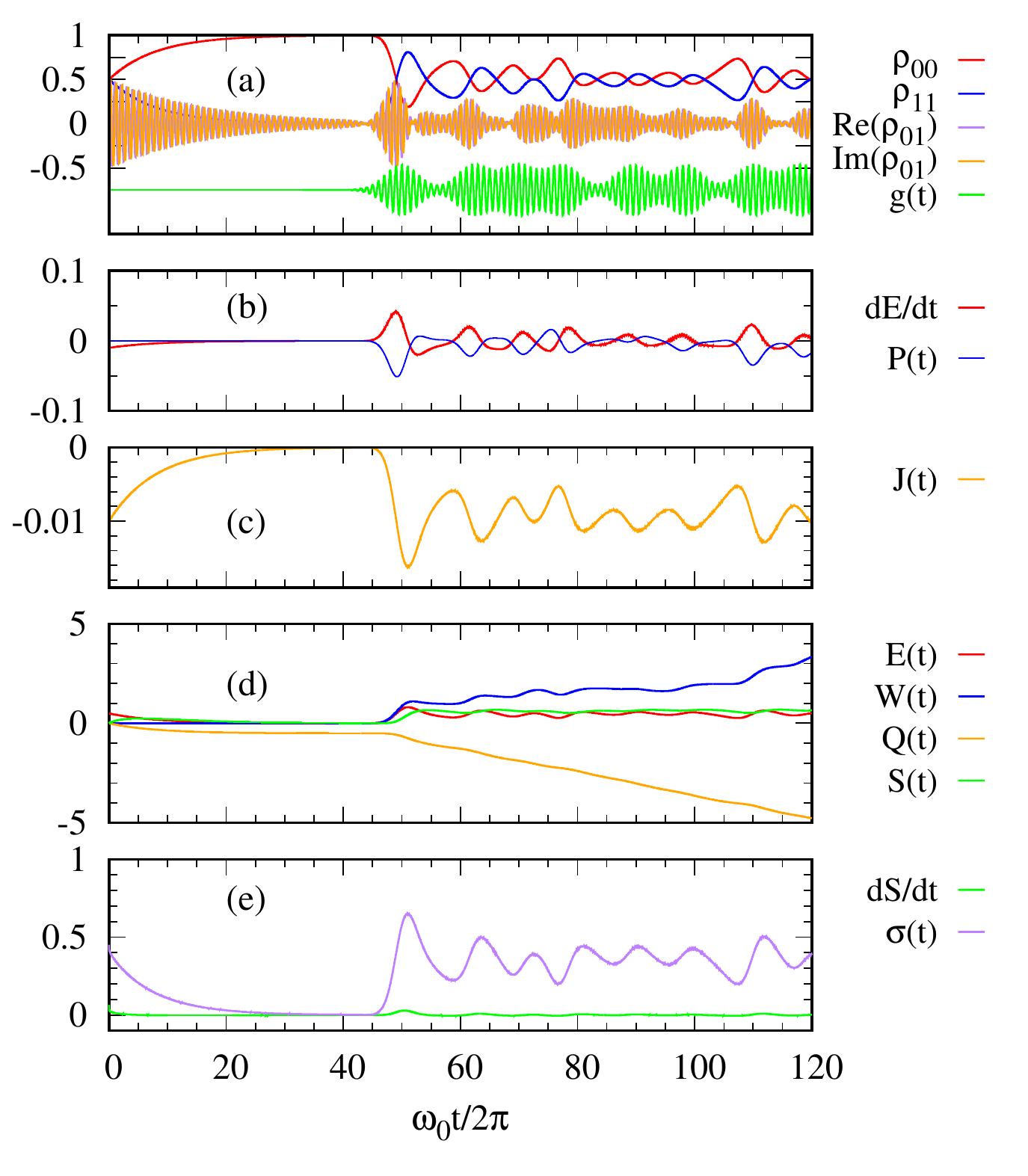}
}
\caption{(a) The density matrix elements of the two-level system and the sequence of Gaussian photon 
pulses $g(t)$,  (b) rate of energy change $\frac{d E(t)}{dt}$ and power $P(t)$, (c) heat current $J(t)$,  
(d) energy $E(t)$, work $W(t)$, heat $Q(t)$ and system entropy $S(t)$, (e) rate of system entropy change and 
entropy production, all as functions of time. This describes the case of irregularly spaced sequence of photon pulses.
Parameters: $\langle n\rangle  = 1$, $\gamma = 10^{-2}\omega_0$, $\Omega = \omega_0/4\pi$,
and $\hbar\omega_0 = 1{\;\rm eV}$.}
\label{Fig5}
\end{figure}

With regard to the first law of quantum thermodynamics, Fig.~\ref{Fig2} (b) and (c) plot the rate of 
energy change, the power and the heat current. The heat current $J(t)$ from the environment to the system
is always negative. This means the excited state of the two-level system releases its energy to the cold bath. 
In contrast, the power $P(t)$ and $\dot{E}(t)$ oscillates out of phase while the photon pulse is applied. 
Fig.~\ref{Fig2} (d) shows the entropy $S(t)$ of the two-level system together with its energy $E(t)$, 
the work done $W(t)$ on the system and the heat transfer $Q(t)$.  Fig.~\ref{Fig2} (e) shows the entropy 
production $\sigma(t) = \dot{S}(t) - \beta \dot{Q}$ as a function of time in relation to the second law 
of thermodynamics. The entropy production $\sigma(t)$ is always positive confirming the second law.

In Fig.~\ref{Fig3}, we see more oscillatory behaviors in the quantities due to a stronger photon pulse with 
$\langle n\rangle  = 10$ than those with $\langle n\rangle  = 1$ in Fig.~\ref{Fig2}. 
Nevertheless, the overall trend is similar to that explained above for Fig.~\ref{Fig2}.

We now examine how the quantum thermodynamic quantities depend on the temporal shape of Gaussian pulse sequence. 
In Figs.~\ref{Fig4} and~\ref{Fig5}, we plot the same quantities as those in Figs.~\ref{Fig2} and~\ref{Fig3}, 
but compare two cases, that is, regularly spaced (Fig.~\ref{Fig4}) and irregularly spaced (Fig.~\ref{Fig5}) 
sequence of Gaussian pulses with the same mean number $\langle n\rangle =|\alpha|^2=1$. As described by 
the curve $g(t)$, the peak times $t_i$'s in Eq.~(\ref{Eq:Gaussian_pulse}) are regularly (not regularly) 
spaced in the left (right) panel. 
In both cases, the overall trend of the thermodynamic quantities are similar to that explained for Fig.~\ref{Fig2}
while the actual response of the system does depend on the temporal shape of the pulse sequence. Remarkably, 
we see that the output power $P(t)$ (blue curve in (b)) and the accumulated work $W(t)$ (blue curve in (d)) 
depend on the temporal shape of the incoming pulses even with the same $|\alpha|^2=1$, which can have implications 
for practical photocell operation. In particular, we find that the case of regular sequencing of pulses yield 
a higher value of work.

\section{Quantum photocell driven by photon pulses}
\label{Sec:Photocell}

\begin{figure}[t]
\centering{
\includegraphics[width=0.35\textwidth]{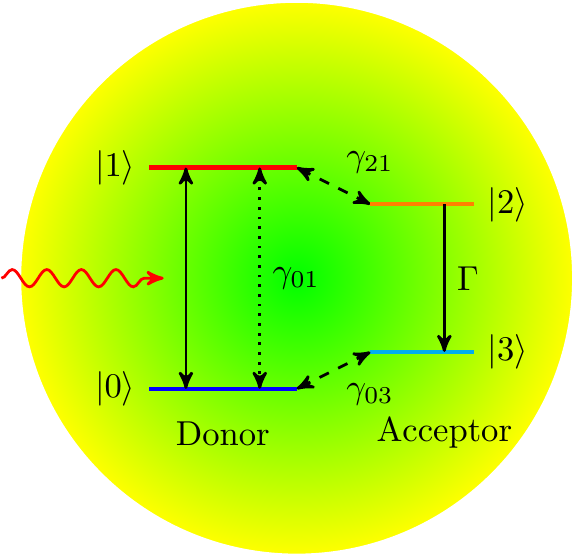}
}
\caption{Schematic diagram of a donor-acceptor photocell. $\gamma_{01}$ is the spontaneous decay 
due the coupling with the cold thermal bath. $\gamma_{21}$ and $\gamma_{03}$ are the transfer 
rate between the donor and the acceptor. $\Gamma$ stands for the external load or 
an electrical resistance.}
\label{Fig:Qphotocell}
\end{figure}

For a quantum heat engine, let us now consider a quantum photovoltaics cell driven by photon pulses, 
as shown in Fig~\ref{Fig:Qphotocell}. The quantum photocell we consider is a 4-level quantum system composed 
of a donor and an acceptor. In 1959, Scovil and Du-Bois~\cite{Scovil59} consider the 3-level system as 
the simplest quantum heat engine where one part of the 3-level system is in thermal equilibrium with
a hot bath and the other part with a cold bath. Many previous studies~\cite{Scully2010,Scully2011,
Svidzinsky2011,Creatore2013,Oh2015,Fruchtman2016,Higgins2017,Dorfman2013,Killoran2015} took a similar 
assumption that the donor of the quantum photocell is in contact with the hot bath, i.e., the sun, and
the acceptor is in thermal contact with the cold bath. In contrast to the previous works, we assume
that the quantum photocell is in thermal contact only with the cold bath.
In our previous work~\cite{Oh2019}, the pumping term was introduced in the Lindblad master equation 
to describe the energy flow from the hot bath. Here the input energy is supplied by the sequence of incoming 
photon pulses. 

The cyclic operation of the quantum photocell can be performed with the sequence as 
follows: (i) The donor absorbs incoming photons and the electron becomes excited with the transition from 
the ground state $\ket{0}$ to the excited state $\ket{1}$. (ii) The phonon vibration makes the excited 
electron at the donor transfer to the acceptor state $\ket{2}$. (iii) The acceptor is coupled to an external 
load and the current flow (electric work) is represented by the transition decay from the state $\ket{2}$ 
to the state $\ket{3}$. (iv) The electron in the state $\ket{3}$ of the acceptor returns to 
the ground state $\ket{0}$ of the donor by a vibrational or non-radiative decay. 

The unperturbed Hamiltonian of the quantum photocell with 4-levels is written as
\begin{equation}
H_0 = -E_0\ketbra{0}{0} -E_1\ketbra{1}{1} -E_{2}\ketbra{2}{2} -E_{3}\ketbra{3}{3}\,.
\label{Eq:photocell_H0}
\end{equation}
Similar to Eq.~(\ref{Eq:JC_Hamiltonian}), the interaction of the donor of the photocell with the incoming 
photon pulses is again described by the Jaynes-Cummings Hamiltonian 
\begin{equation}
H_1(t) = i\hbar\left[ g^*(t)\sqrt{\gamma}\,\sigma_{-} - g(t)\sqrt{\gamma}\,\sigma_+ \right]\,,
\label{Eq:photocell_H1}
\end{equation}
where $\sigma_+ = \ketbra{1}{0}$ and $\sigma_+ = \ketbra{0}{1}$.
Same as Eq.~(\ref{Eq:two_level_Dc}), the interaction of the donor of the quantum photocell with 
the cold thermal bath is represented by the Lindblad operator,
\begin{align}
{\cal D}_c[\rho]
&= \frac{\gamma_{01}}{2}\bigl(\bar{n}_c + 1\bigr)
    \bigl(2\sigma_{-}\rho\sigma_{+} -\sigma_{+}\sigma_{-}\rho + \rho\sigma_{+}\sigma_{-}\bigr) \nonumber\\
&+\frac{\gamma_{01}}{2}\bar{n}_c
\left(2\sigma_{+}\rho\sigma_{-} -\sigma_{-}\sigma_{+}\rho + \rho\sigma_{-}\sigma_{+}\right)\,.
\label{Eq:photocell_Dc}
\end{align}

The electron transfer between the states $\ket{1}$ and $\ket{2}$ and that 
between the state $\ket{3}$ and $\ket{0}$ are described by the Lindblad operator 
${\cal D}_{\rm ph}[\rho] = {\cal D}_{\rm ph}^{(1,2)}[\rho] + {\cal D}_{\rm ph}^{(3,0)}[\rho]$, 
where 
\begin{align}
{\cal D}_{\rm ph}^{(i,j)}[\rho]
&= \frac{\gamma_{ij}}{2}\bigl(\bar{n}_{\rm ph} + 1\bigr)
    \bigl(2L_{ij}\rho L^\dag_{ij} -L^\dag_{ij}L_{ij}\rho + \rho L^\dag_{ij}L_{ij} \bigr) \nonumber\\
&+\frac{\gamma_{ij}}{2}\bar{n}_{\rm ph}
\left(2L^\dag_{ij}\rho L_{ij} -L_{ij}L^\dag_{ij}\rho + \rho L_{ij}L^\dag_{ij}\right)\,.
\label{Eq:photocell_Dij}
\end{align}
Here $\gamma_{12}$ and $\gamma_{30}$ represent the transition rate between $\ket{1}$ and $\ket{2}$
and between $\ket{3}$ and $\ket{0}$, respectively.  $L_{ij} = \ketbra{i}{j}$ and 
$L_{ij}^\dag = \ketbra{j}{i}$ are the lowering and raising operators, respectively. 
$\bar{n}_{\rm ph}$ is the phonon occupation number at $\hbar\omega =E_1-E_2=E_3-E_0$ and $T_c=300\;{\rm K}$.
The work done by the quantum photocell to the external load is described by the ohmic dissipation 
\begin{equation}
{\cal D}_{\rm ohm}[\rho]
= \frac{\Gamma}{2}\Bigl(2L_3\rho L_3^\dag -L_3^\dagger L_3\rho +
\rho L^\dagger_3 L_3\Bigr)\,,
\label{Eq:photocell_ohm}
\end{equation}
where $L_3 = \ketbra{3}{2}$. Here $\Gamma$ represent the conductance of the external load and may 
be changed from zero corresponding to the open circuit and to a big number representing 
the short-circuit of the quantum photocell. With 
Eqs.~(\ref{Eq:photocell_H0}), (\ref{Eq:photocell_H1}), (\ref{Eq:photocell_Dc}), 
(\ref{Eq:photocell_Dij}), (\ref{Eq:photocell_ohm}), the LGKS equation for the quantum photocell is 
written as 
\begin{align}
\frac{d}{dt}\rho(t) 
&= -\frac{i}{\hbar}[H_0 + H_1(t),\rho(t)] + {\cal D}_c[\rho] \nonumber\\
&+ {\cal D}_{\rm ph}^{(1,2)}[\rho] + {\cal D}_{\rm ph}^{(3,0)}[\rho] + {\cal D}_{\rm ohm}[\rho]\,.
\label{Eq:Lindblad_photocell}
\end{align}

Since the quantum photocell has no direct interaction Hamiltonian between the donor and the acceptor, 
the Hamiltonian $H_S(t) = H_0 + H_1(t)$ of the quantum photocell can be written as the sum of the 
time-dependent donor Hamiltonian $H_D$ and the time-independent acceptor Hamiltonian $H_A$,
\begin{align}
H_S(t) = H_D(t) + H_A\,,
\end{align}
where $H_D(t) \equiv -E_0\ketbra{0}{0} -E_1\ketbra{1}{1} + H_1(t)$ and 
$H_A \equiv -E_2\ketbra{2}{2} - E_3\ketbra{3}{3}$.
This partition makes it possible to express some quantum thermodynamic quantities as the sum of 
the donor and acceptor parts. The energy of the quantum photocell is given by
the sum of the energies of the donor and acceptor
\begin{align}
E(t) = {\rm tr}\left\{\rho(t) H_S(t)\right\} = E_D(t) + E_A(t) \,,
\end{align}
where the donor energy $E_D(t)$ and the acceptor energy $E_A(t)$ are given by  
\begin{subequations}
\begin{align}
E_D(t) &= {\rm tr}_D\{ \rho_D(t) H_D(t)\} \,, \\
E_A(t) &= {\rm tr}_A\{ \rho_A(t) H_A\},
\end{align}
\end{subequations}
respectively. Here $\rho_{A} = {\rm tr}_{D}\{\rho\}$ and $\rho_{D} = {\rm tr}_{A}\{\rho\}$ are the density 
operators of the donor and acceptor, respectively. Since the photon pulse delivers the power only to 
the donor, the power $P(t)$ is given by the power of the donor
\begin{align}
P(t) = {\rm tr}\{\rho(t)\dot{H}_S(t)\} 
     = {\rm tr}_D\{\rho_D \dot{H}_1(t)\} \equiv P_D(t)\,.
\label{Eq:power_photon}
\end{align}
The heat dissipation occurs at the donor and acceptor. Thus the heat current $J(t)$ is written as
the sum of the two parts
\begin{align}
J(t) &= {\rm tr}_D\{\dot{\rho}_D(t)H_D(t)\} + {\rm tr}_A\{\dot{\rho}_A(t)H_A\}\nonumber\\ 
     &= J_D(t) + J_A(t)\,.
\end{align}
Here the heat current of the acceptor $J_A$ may be associated with the power delivered by the quantum photocell 
to the external load. Finally, the entropy of the quantum photocell $S(t)={\rm tr}\{\rho\log\rho\}$
can be written as the sum of the entropies of the donor and acceptor, 
$S_D(t) = -{\rm tr}_D\{\rho_D\log\rho_D\}$ and $S_A(t) =- {\rm tr}_A\{\rho_A\log\rho_A\}$, too. This is because 
there is no coherent interaction between the donor and the acceptor so the whole density operator of the system 
has a structure $\rho=\rho_D\oplus \rho_A$. 

\begin{figure}[t]
\centering{
\includegraphics[width=0.5\textwidth]{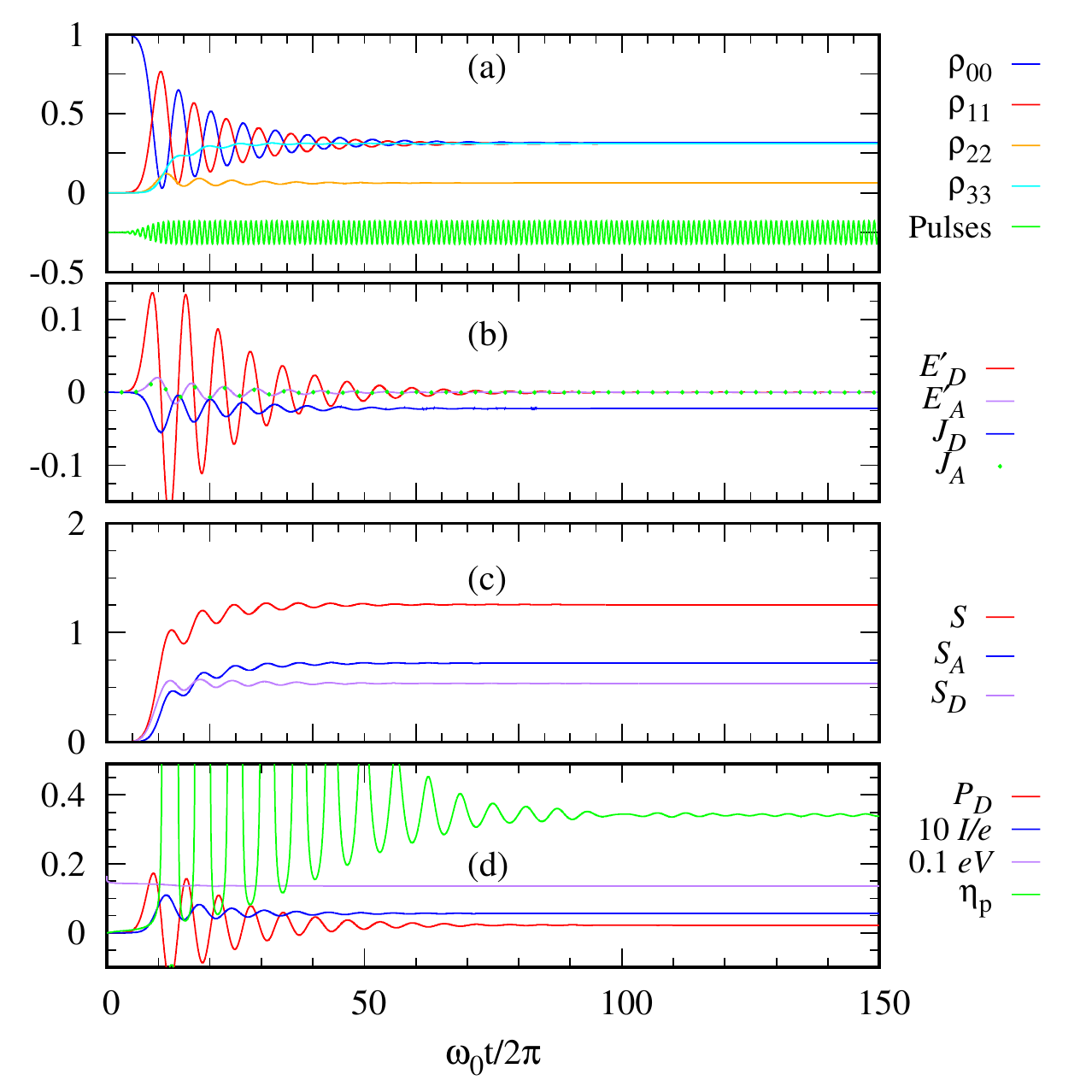}
}
\caption{(a) The diagonal matrix elements of the density operator of the photocell and the pulse profiles
are plotted as a function of time. (b) The changes in the energy of the donor and acceptor, $({dE}/{dt})_D = E'_D$
and $({dE}/{dt})_A = E_A'$, the heat currents of the donor and acceptor,$J_D$ and $J_A$, are plotted as
a function of time. (c) The entropy of the quantum photocell, $S(t)$, the entropy of the donor, $S_D(t)$, and
the entropy of the acceptor, $S_A(t)$, are calculated as a function of time. The power $P_D(t)$ delivered to the donor
by the photon pulses, the current $I(t)$, the voltage $V(t)$, and the efficiency $\eta$ are plotted
as a function of time.  Parameters: $\langle n\rangle  = 10$, $\gamma = 10^{-3}\omega_0$, $\gamma = 10^{-2}\omega_0$ 
$\Gamma = 0.1\omega_0$, $\Omega = \omega_0/4\pi$, and $\hbar\omega_0 = E_1 -E_0 = 1.8\;{\rm eV}$.
}
\label{Fig7}
\end{figure}

We calculate the current through the external load as 
\begin{equation}
I = e \Gamma\cdot \rho_{22},
\end{equation}
and the voltage across the external load 
\begin{equation}
eV = E_2 -E_3 + k_BT\log\left(\frac{\rho_{22}}{\rho_{33}}\right).
\end{equation}
The latter comes from the relation $\rho_{22}/\rho_{33} =\exp(-(E_2 -E_3 -eV)/k_BT)$. 
The electric power delivered to the external load by the photocell is written as 
$P_{\rm out} = I(t)\cdot V(t)$ which depends on the external conductance $\Gamma$.
Now that we have the power delivered by the photon pulse, Eq.~(\ref{Eq:power_photon}) and the electric
power output $P_{\rm out}(t)$, we can define the power efficiency of the quantum photocell as
\begin{equation}
\eta_p = \frac{P_{\rm out}(t)}{P_D(t)}\,.
\end{equation}

\begin{figure}[t]
\begin{center}
\includegraphics[width=0.5\textwidth]{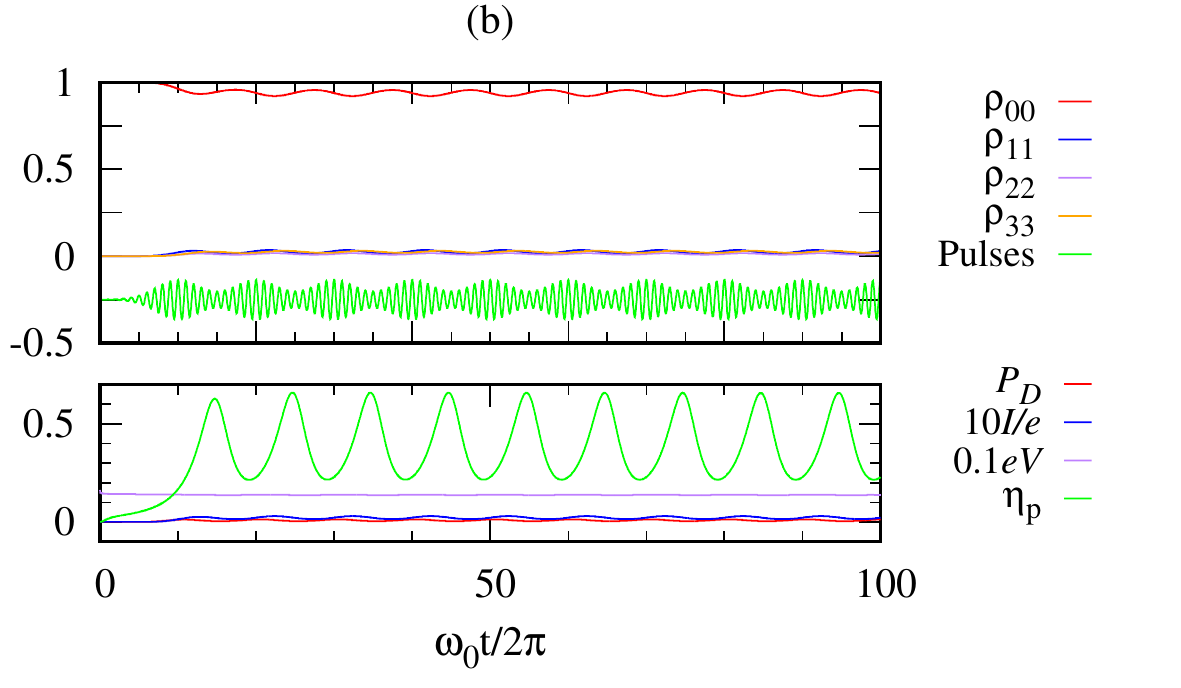}\\
\includegraphics[width=0.5\textwidth]{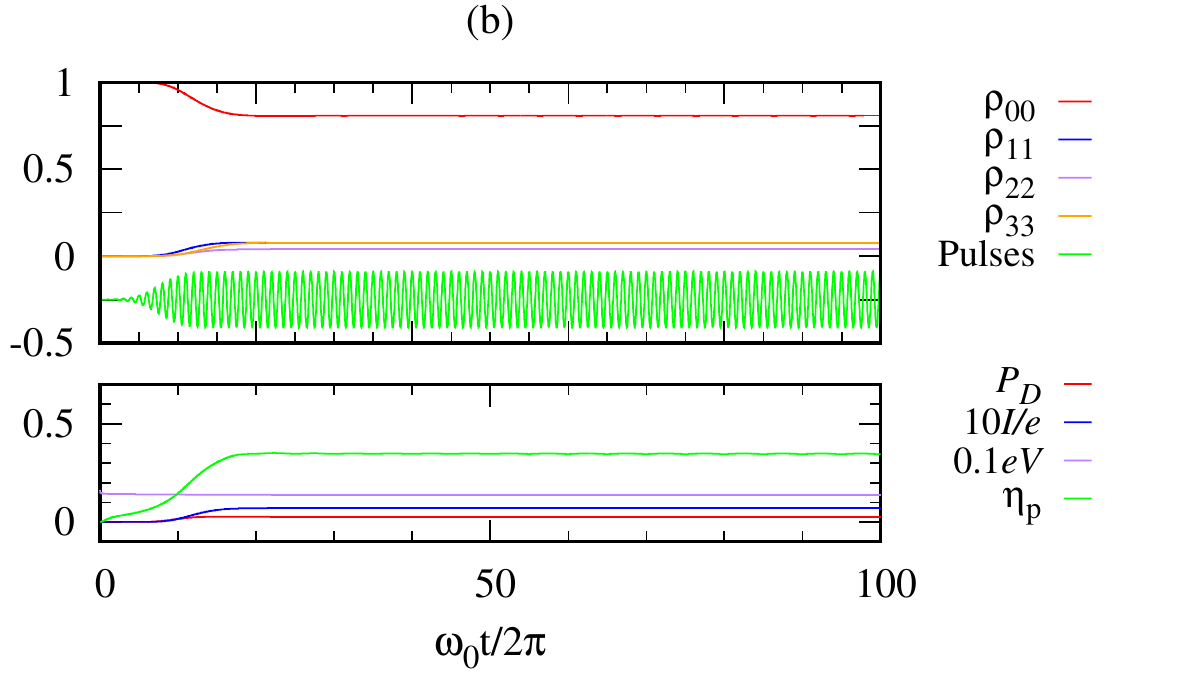}
\end{center}
\caption{From discrete mode to the continuous mode operation  by changing the interval of the pulses.
Parameters: $\langle n\rangle  = 1$, $\gamma = 10^{-3}\omega_0$, $\gamma = 10^{-2}\omega_0$ 
and $\Omega = \omega_0/4\pi$.}
\label{Fig6}
\end{figure}

We solve numerically the LGKS Eq.~(\ref{Eq:Lindblad_photocell}) using the Runge-Kutta method for 
different sequences of photon pulses to obtain the quantum thermodynamic quantities.
Fig.~\ref{Fig7} (a) plots the population of each level of the quantum photocell over time 
when the sequences of the Gaussian pulses are applied one immediately after another almost in a continuum limit. 
Fig.~\ref{Fig7} (b) shows the power $P_D(t)$ delivered to the donor by the photon, the heat dissipation at 
the donor and the acceptor, $J_D(t)$ and $J_A(t)$, the change in energies of the donor and acceptor,
$\dot{E}_D(t)$ and $\dot{E}_A(t)$. Fig.~\ref{Fig7} (c) show the total entropy $S(t)$ of the system
and the entropies of the donor and the acceptor, $S_D(t)$ and $S_A(t)$. Our numerical calculation
confirms that the total entropy is the sum of those of the donor and the acceptor, $S(t) = S_D(t) + S_A(t)$, 
as explained before.

Fig.~\ref{Fig7} (d) depicts the the current $I(t)$, the voltage $V(t)$, the electric power output $P_{\rm out}(t)$, 
the power delivered by the photon $P_D(t)$, and the power efficiency $\eta$. From the figures, we see that these 
quantities initially show an oscillatory behavior then become saturated in the long time limit. In particular, 
the asymptotic value of power efficiency is as high as $\eta_p\sim 0.36$, which can also be interpreted as the work 
efficiency, i.e. work output divided by energy input, when $\eta_p$ is constant. 

There are different types of quantum heat engines like continuous engine, two-stroke engine and four-stroke engine. 
Many other studies considered the quantum photocell as the continuous heat engine where the donor is in thermal 
contact with the hot reservoir and the acceptor is in the cold bath. Fig.~\ref{Fig7} demonstrates the photocell as 
a continuous heat engine, which is not in contact with a hot bath, but is supplied input energy by photon pulses. 
In our case, we have a flexibility of engineering the input photon pulses as desired.  In Fig.~\ref{Fig8}, we further 
compare the power efficiency between two cases. Fig.~\ref{Fig8} (a) is the case where the photon pulses are applied 
at a finite time interval (discrete mode operation). On the other hand, Fig.~\ref{Fig8} (b) is the case where 
the photon pulses are applied almost continuously. Both have the same energy parameter $\langle n\rangle = 1$ of 
incoming Gaussian pulses. In the discrete mode, we see an oscillatory behavior of power efficiency between 0.2 
and 0.6. In the continuum mode, the power efficiency does not oscillate but asymptotically approaches the value 
$\eta_p\sim0.36$. 

Let us compare the two cases in Fig.~\ref{Fig7} and Fig.~\ref{Fig8} (b) both dealing with the continuum limit with 
different energies $\langle n\rangle  = 10$ and $\langle n\rangle  = 1$, respectively. It is interesting to find 
that the efficiency turns out to be the same $\eta_p\sim0.36$ regardless of the pulse energy of our consideration. 
Of course, the transient behaviors are different as the high-energy case shows an oscillatory behavior while the 
low-energy case does not. Aside from details, we see that our model of heat engine offers a possibility to make 
an efficient quantum engine with a proper design.

\medskip
\section{Summary}
\label{Sec:Summary}

In this paper, we studied quantum thermodynamics of two open quantum systems, the two-level system and 
the quantum photovoltaic model, driven by the Gaussian photon pulses. By solving the master equation
with the time-dependent Hamiltonian of the Gaussian photon pulses, we calculated quantum thermodynamic 
quantities. For the two-level system in the cold bath, we examined the first law of quantum thermodynamics, 
which relates the energy change of the system, the heat current, and the power. 
We also illustrated the second law of thermodynamics by confirming that the entropy production is positive.

More importantly, we investigated the quantum photovoltaic cell in the cold bath driven by the sequence of the Gaussian 
photon pulses. The power efficiency of the quantum 
photocell was considered as the ratio of the output power delivered 
to the external load by the photocell to the input power delivered by the photon pulses. 
We showed that the quantum photocell as a heat engine can operate both in the discrete stroke mode and in the continuous stroke mode by changing the sequence of 
the photon pulses. 

Our model of quantum heat engine based on a driven quantum system in contact with a single bath seems worthwhile to further investigate. In our work we showed that an efficiency as high as $\eta_p=0.36$ can be achieved, which should be further explored in a broad range of system parameters. There are some meaningful directions to consider. One is to study 
how the dark state or the quantum coherence can further enhance the performance of the photocell. 
We also note that recently Chan {\it et al.}~\cite{Chan2018} 
studied the quantum dynamics of excitons by absorption of single photons in photosynthetic 
light-harvesting complexes. It would be interesting how the photosynthetic light-harvesting complexes 
behave when the photon pulses are applied. Moreover, while we considered the Gaussian photon pulses in the current work, other photon pulses, 
for example, hyperbolic secant, rectangular, or symmetric exponential pulses may be tested to come up with an optimal design ~\cite{Wang2011}.
An open problem is how to mimic the thermal photon from the hot thermal bath and to incorporate the thermal
photons into the simulation. The quantum photocells considered here can be simulated on quantum 
computers~\cite{Potocnik2018,Maslennikov2019}.

\bibliography{Q_driven_manuscript_v1.bib}
\end{document}